\documentclass[aps,prl,superscriptaddress,twocolumn,10pt,notitlepage,longbibliography]{revtex4-1}

\usepackage[utf8]{inputenc}
\usepackage[T1]{fontenc} 
\usepackage{float}
\usepackage{braket}
\usepackage{amsmath,amsfonts,amssymb,eqnarray}  
\usepackage{amsthm, color, soul}
\usepackage{graphicx,subfigure}  
\usepackage{array,booktabs}
\usepackage[pdftex,hypertexnames=false]{hyperref} \usepackage[capitalize]{cleveref}
\usepackage{multirow}

\begin{document}
\title{Orbital angular momentum states enabling fiber-based high-dimensional quantum communication} 

\author{Daniele Cozzolino}
\email{dacoz@fotonik.du.dk}
\author{Davide Bacco}
\email{dabac@fotonik.dtu.dk}
\author{Beatrice Da Lio}
\author{Kasper Ingerslev}
\author{Yunhong Ding}
\author{Kjeld Dalgaard}
\affiliation{CoE SPOC, Dep. Photonics Eng., Technical University of Denmark, Kgs. Lyngby 2800, Denmark}
\author{Poul Kristensen}
\affiliation{OFS-Fitel, Priorparken 680, DK-2605, Broendby, Denmark}
\author{Michael Galili}
\author{Karsten Rottwitt}
\affiliation{CoE SPOC, Dep. Photonics Eng., Technical University of Denmark, Kgs. Lyngby 2800, Denmark}
\author{Siddharth Ramachandran}
\affiliation{Electrical and Computer Engineering Department, Boston University, 8 St Mary's St, Boston, MA, USA}
\author{Leif Katsuo Oxenl\o we}
\affiliation{CoE SPOC, Dep. Photonics Eng., Technical University of Denmark, Kgs. Lyngby 2800, Denmark}

\date{\today}

\begin{abstract}
\noindent
Quantum networks are the ultimate target in quantum communication, where many connected users can share information carried by quantum systems. The keystones of such structures are the reliable generation, transmission and manipulation of quantum states. Two-dimensional quantum states, qubits, are steadily adopted as information units. However, high-dimensional quantum states, qudits, constitute a richer resource for future quantum networks, exceeding the limitations imposed by the ubiquitous qubits. The generation and manipulation of such $D$-level systems have been improved over the last ten years, but their reliable transmission between remote locations remains the main challenge.
Here, we show how a recent air-core fiber supporting orbital angular momentum (OAM) modes can be exploited to faithfully transmit $D$-dimensional states. Four OAM quantum states and their superpositions are created, propagated in a 1.2 km long fiber and detected with high fidelities. In addition, three quantum key distribution (QKD) protocols are implemented as concrete applications to assert the practicality of our results. This experiment enhances the distribution of high-dimensional quantum states, attesting the orbital angular momentum as vessel for the future quantum network.
\end{abstract}

\maketitle

\section{Introduction}
Quantum communication, \textit{i.e.} the faithful transmission of quantum states between separated users, has established itself in the last decade due to its inherently and fascinating connection to quantum nonlocality and ultimate communication security~\cite{GisinQC}. The distribution of quantum states has been investigated by using different links between parties, such as satellites~\cite{Ren2017,Yin2017}, free-space~\cite{Stein2017,Sit2017}, fibers~\cite{Frohlich2017,Korzh2015} and very recently underwater links~\cite{Bouchard_under,Ling2017}. The majority of quantum communication experiments use two-level quantum systems (qubits) as fundamental information units. Nevertheless, high-dimensional quantum states (qudits) offer more practical advantages in quantum communication and information in general. For instance, high-dimensional quantum states enhance violation of Bell-type inequality tests of local realism~\cite{Kaszlikowski2000,Durt2001,Collins2002,Vertesi2010,Dada2011}, they provide efficiency improvements to communication complexity problems~\cite{Brukner2002,Brukner2004} and constitute richer resources for quantum computation ~\cite{Lanyon2009,Babazadeh2017,Muralidharan2017} and metrology~\cite{Fickler2012}. Specifically for communication purposes, qudits increase the information content per photon~\cite{Fickler2012,BP2000}, they provide higher noise resistance~\cite{Cerf2002} and improve security for quantum key distribution (QKD) protocols~\cite{Durt2004,Erhard2018}. To date, high-dimensional quantum states have been implemented by exploiting position-momentum~\cite{Edgar2012,Walborn2006,Etcheverry2013} and orbital angular momentum (OAM)~\cite{Malik2016,Mirhosseini2015,Anton2006} degrees of freedom, spatial modes in multicore fibres~\cite{Ding2016,Lee2017,Canas2016}, time-energy and time-bin encoding~\cite{Zhong2015,Alikhan2007,IslamTimeBin2017}.
Furthermore, great steps have been made to generate and manipulate quantum states of higher dimensions~\cite{Wang2018,kues2017,Bavaresco2018}, however their transmission, the cornerstone for the future quantum network, has not been exhaustively investigated. Indeed, quantum communication usefulness and range of applicability strongly depends on the extent to which quantum channels can be reliably and efficiently implemented. In particular, their transmission over a safe enclosed environment like a fiber has so far not been demonstrated in an unbounded information preserving manner. Standard single-mode fibers have been extensively used for time-bin qudit distribution, since those states are preserved throughout the transmission~\cite{Zhong2015,IslamTimeBin2017}. Nonetheless, the main drawback of this high-dimensional communication relies on the encoding scheme. Indeed, increasing the dimensions in the time-energy encoding, with the reasonable assumption of fixed repetition rate at the transmitter, the detected photon rate is lowered, as well as the overall information. Position-momentum high-dimensional quantum states have been transmitted through standard multimode fibers~\cite{Walborn2006}. However, due to the high crosstalk between the optical modes during the propagation, the information is not preserved, making standard multimode fibers inefficient for qudit distribution. Finally, multicore fibers have been utilized to transfer spatial mode of light~\cite{Ding2016,Canas2016}, but phase drifts among cores impair coherence of high-dimensional states.
\begin{figure}[h!]
   \centering
    {\includegraphics[width=8.3cm]{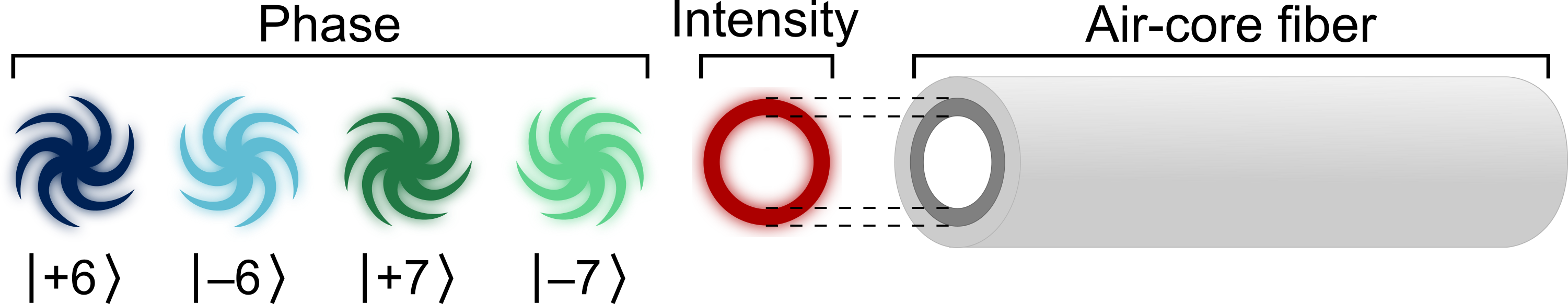}}
    \caption{ \textbf{Schematic of the transmission of OAM modes.} The picture shows the four different OAM states implemented in this work. Their spiraling phases with different orientation and intertwined number of helices are shown as well. The rotating phase characterizing the OAM modes generates a singularity along the beam axis, which results in a ring-shaped intensity distribution. The air core fiber structure is particularly suitable to carry OAM modes, avoiding mode mixing and thus preserving their orthogonality.}
\label{fig:scheme}
\end{figure}
Fiber transmission of qudits based on the OAM, the emblematic degree of freedom to generate high-dimensional quantum states, has not been proven yet, but classical and quantum communication with qubits have been demonstrated so far~\cite{Bozinovic2013,Kasper,Sit2018,BozinovicFio}.\\
Here, we prove the first transmission of high-dimensional quantum states, encoded using their orbital angular momentum, by exploiting a recently developed air-core fiber~\cite{Gregg2015} which transcend the constraints of previous approaches. Indeed, it presents an enhanced preservation of mode orthogonality, which results in a negligible mode mixing between the OAM modes simultaneously propagated. As schematically shown in figure~\ref{fig:scheme}, four different OAM states and their superpositions are generated, transmitted and detected, certifying the feasibility of high-dimensional quantum communication using this fiber. To emphasize the practicality of our result, we implement two-dimensional (2$D$), four-dimensional (4$D$) and two times ($2\times$) multiplexed 2$D$ QKD protocols and compare their performances. Our results settle the orbital angular momentum as one of the most promising degrees of freedom to distribute high dimensional quantum states in a future quantum network. 

\section{Fiber-based distribution of high-dimensional quantum states}
\begin{figure*}[ht!]
   \centering
    {\includegraphics[width=17.3cm]{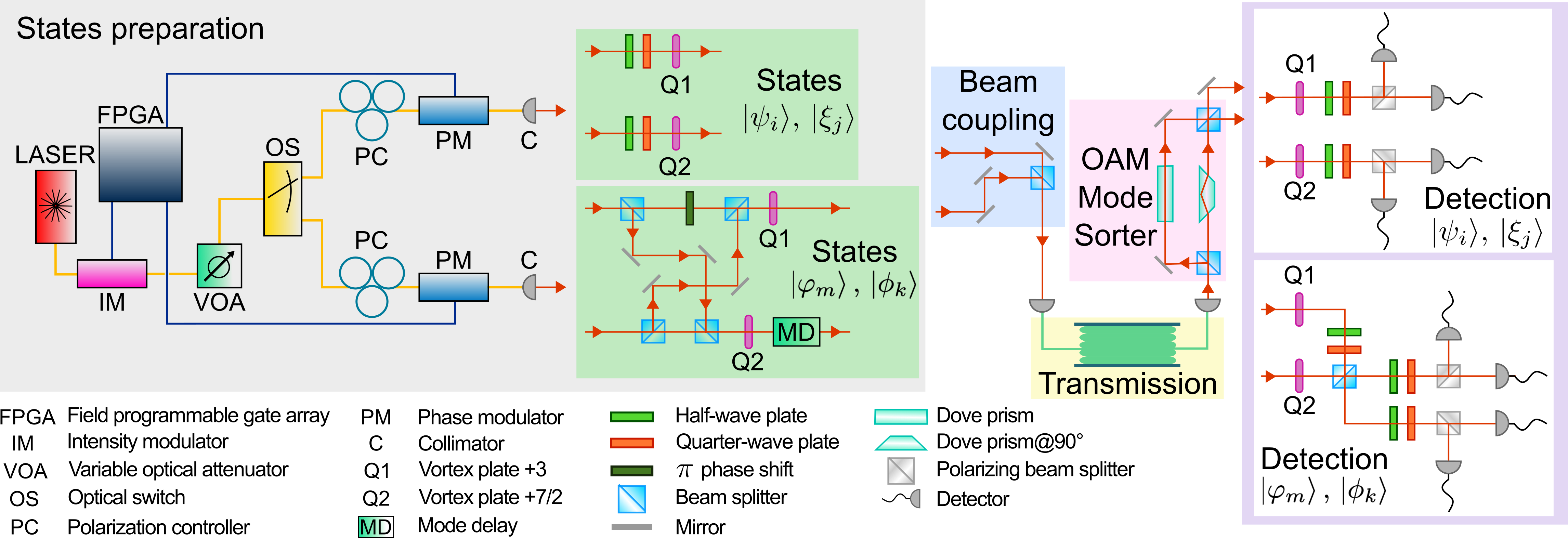}}
    \caption{ \textbf{Experimental setup.} A continuous-wave laser ($1550$ nm) is carved into pulses by an intensity modulator (IM) driven by a field programmable gate array (FPGA). A variable optical attenuator (VOA) allows the quantum regime to be reached. The polarization of the quantum states is prepared by phase modulators (PMs). Two different voltages are chosen to yield diagonal (D) and antidiagonal (A) polarizations at the output of the PMs. Half- and quarter-wave plates are used to convert the linear to circular polarization when needed. Vortex plates with topological charge $q_1=+3$ and $q_2=+7/2$ (Q1, Q2) assign different OAM quantum number to the incoming photons. The quantum states are then coupled into the air-core fiber. After the fiber transmission, an OAM mode sorter is implemented to separate photons with even OAM from those with odd OAM modes ($\ell=|6|$ and $\ell=|7|$). Two other vortex plates, identical to the previous, are used to convert from OAM to Gaussian-like modes. After the OAM sorter, the quantum states $\ket{\psi_i}$, $\ket{\xi_j}$, $\ket{\varphi_m}$ and $\ket{\phi_k}$ are detected using their polarization. Other sets of half- and quarter-wave plates are used to rotate the polarization to be linear, while a polarization beam splitter (PBS) is used to separate between its two orthogonal components. To measure the relative phase between different angular momenta ($|\ell|$-values) in the $\ket{\varphi_m}$ and $\ket{\phi_k}$ states, a free-space Mach-Zehnder interferometer is used at the output of the mode sorter. Four single-photon detectors are connected to the outputs. The insertion loss is measured to be approximately $9$ dB for the $\ket{\psi_i}$ and $\ket{\xi_j}$  receiver and $10$ dB for the $\ket{\varphi_m}$, $\ket{\phi_k}$ receiver.}
\label{fig:mainsetup}
\end{figure*}

\begin{figure*}[ht!]
     \centering
     \includegraphics[width=1\linewidth]{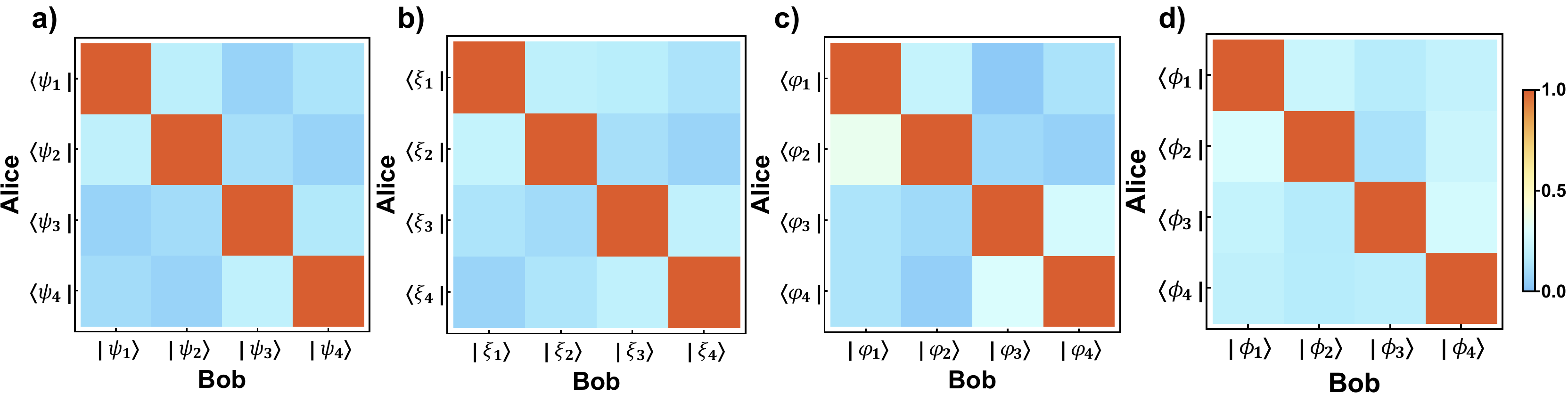}
   \caption{ \textbf{Experimental data certifying the purity of the transmitted states.} \textbf{a)}--\textbf{d)} The states $\ket{\psi_i}$, $\ket{\xi_j}$, $\ket{\varphi_m}$ and $\ket{\phi_k}$ respectively implemented, transmitted and detected. For each set of states, one state per time is prepared while all the possible outcomes are detected simultaneously. Table \ref{table:fidelity} reports the obtained fidelities for each set of states.}
\label{fig:tomos}
\end{figure*}

Photons carrying orbital angular momentum are characterized by a helical phase factor $e^{i\ell\theta}$, where $\theta$ is the azimuthal coordinate and $\ell$ is an unbounded integer value representing the quanta of OAM that each photon possesses~\cite{Allen1992}. Different values of $\ell$ represent discrete states on which a high-dimensional basis set can be devised. Table~\ref{table:states} shows all the states prepared and detected for these experiments. The experimental implementation is shown in figure~\ref{fig:mainsetup}. Weak coherent pulses, with OAM quantum number $\ell\in\lbrace -7,-6,+6,+7 \rbrace$, are prepared with two vortex plates having topological charge $q_1=+3$ and $q_2=+7/2$ (see Appendix A)~\cite{qplate}. Leveraging on the spin-orbit coupling, we can generate the OAM quantum states $\ket{-6}, \ket{+6}, \ket{-7}$, $\ket{+7}$ by swapping between left- and right-handed (L, R) circular polarizations, \textit{i.e.} by flipping the spin angular momentum (SAM) of the impinging photons. Indeed, an L-polarized beam acquires a $+\ell$ OAM charge with R polarization when passing through the vortex plate. Like wise, an R-polarized beam yields a $-\ell$ OAM charge with L polarization. Diagonal and antidiagonal polarizations are used to generate the superposition among states with the same $|\ell|$ order, while superpositions of quantum states with different $|\ell|$ are obtained by combining the states $\ket{|6|}$ and $\ket{|7|}$ with beam-splitters (details in Appendix A). The flipping operation is dynamically regulated by phase modulators followed by sets of suitably oriented half- and quarter-wave plates, yielding the desired L or R polarization. The qudits are transmitted through a 1.2 km-long air-core fiber and detected by an all-optical quantum receiver. The OAM mode sorter depicted in figure~\ref{fig:mainsetup} is based on the scheme proposed by J. Leach et al.~\cite{Leach2002}. It is a Mach–Zehnder interferometer with a Dove prism in each arm, one rotated by $90^\circ$ with respect to the other, which sorts photons based on their OAM quantum number parity. So, depending on the OAM mode of a photon, it leads to constructive or destructive interference. Two detection schemes are implemented to enable detection of all the quantum states prepared. Both receivers present two vortex plates (identical to the previous) to convert back from an OAM mode to a Gaussian-like one, and a combination of polarizing beam-slitters
\begin{table}[h!]
\caption{\textbf{Quantum states implemented.} The table shows all the OAM states generated, transmitted and detected in the experiments. Also listed are the settings for the experimental implementation: the respective polarizations needed to leverage on the spin-orbit coupling, and the phases to create the right superposition states. 
}
\centering
\begin{tabular}{c c c c c c}
\toprule
\multicolumn{2}{c}{\multirow{4}{*}{Quantum states}} &\multicolumn{4}{c}{Exp. implementation}\\
\cline{3-6}
\rule{0pt}{3ex} & & \multirow{2}{*}{Polarization} & \multicolumn{3}{c}{Vortex-plate}\\
\cline{4-6}
\rule{0pt}{3ex} & & & $q_1$ & Phase & $q_2$\\
\hline
\rule{0pt}{3ex}$\ket{\psi_1}$ & $\ket{+6}$ & $L$ & +3 & & \\[1ex]
$\ket{\psi_2}$ & $\ket{-6}$ & $R$ & +3 & & \\[1ex]
$\ket{\psi_3}$ & $\ket{+7}$ & $L$ & & & $+\frac{7}{2}$\\[1ex]
$\ket{\psi_4}$ & $\ket{-7}$& $R$ & & & $+\frac{7}{2}$\\[1ex]
$\ket{\xi_1}$ & $\frac{1}{\sqrt{2}}(\ket{+6} + \ket{-6})$ & $D$ & +3 & & \\[1ex]
$\ket{\xi_2}$ & $\frac{1}{\sqrt{2}}(\ket{+6} - \ket{-6})$ & $A$ & +3 & & \\[1ex]
$\ket{\xi_3}$ & $\frac{1}{\sqrt{2}}(\ket{+7} + \ket{-7})$ & $D$ & & & $+\frac{7}{2}$\\[1ex]
$\ket{\xi_4}$ & $\frac{1}{\sqrt{2}}(\ket{+7} - \ket{-7})$ & $A$ & & & $+\frac{7}{2}$\\[1ex]
$\ket{\varphi_1}$ & $\frac{1}{\sqrt{2}}(\ket{+6} + \ket{+7})$ & $L$ & +3 & + & $+\frac{7}{2}$ \\[1ex]
$\ket{\varphi_2}$ & $\frac{1}{\sqrt{2}}(\ket{+6} - \ket{+7})$ & $L$ & +3 & - & $+\frac{7}{2}$ \\[1ex]
$\ket{\varphi_3}$ & $\frac{1}{\sqrt{2}}(\ket{-6} + \ket{-7})$ & $R$ & +3 & + & $+\frac{7}{2}$ \\[1ex]
$\ket{\varphi_4}$ & $\frac{1}{\sqrt{2}}(\ket{-6} - \ket{-7})$ & $R$ & +3 & - & $+\frac{7}{2}$ \\[1ex]
$\ket{\phi_1}$ & $\frac{1}{2}(\ket{+6} + \ket{-6} + \ket{+7} + \ket{-7})$ & $D$ & +3 & + & $+\frac{7}{2}$\\[1ex]
$\ket{\phi_2}$ & $\frac{1}{2}(\ket{+6} + \ket{-6} - \ket{+7} - \ket{-7})$ & $D$ & +3 & - & $+\frac{7}{2}$\\[1ex]
$\ket{\phi_3}$ & $\frac{1}{2}(\ket{+6} - \ket{-6} + \ket{+7} - \ket{-7})$ & $A$ & +3 & + & $+\frac{7}{2}$\\[1ex]
$\ket{\phi_4}$ & $\frac{1}{2}(\ket{+6} - \ket{-6} - \ket{+7} + \ket{-7})$ & $A$ & +3 & - & $+\frac{7}{2}$\\[0.6ex]
\bottomrule
\bottomrule
\end{tabular}
\label{table:states}
\end{table}
(PBS), half- and quarter-wave plates for SAM analysis. Indeed, by measuring the polarization, we can recover information on the OAM quantum numbers. Specifically for the detection of the superposition quantum states, $\ket{\varphi_m}$ and $\ket{\phi_k}$, after the mode sorter the weak pulses are interfered again to infer the phase relation between two distinct $|\ell|$ OAMs. The detection schemes can register all the possible outcome events at the same time using four single-photon detectors, so that the transmission of the OAM quantum states can be properly characterized. Moreover, such schemes enable complete device-independent demonstrations and loopholes free measurements for nonlocality tests~\cite{Weihs1198,Cyril2011}.More details on the experimental apparatus and the fiber are provided in Appendices B,C and F. To quantify the transmission effects, the purity of the high-dimensional states after the propagation is measured by using the definition of fidelity $F(p,r) = \sum_i (p_i r_i)^{1/2}$, where $p$ and $r$ are the experimental and the theoretical discrete probability distributions of the different set of states with elements $p_i$ and $r_i$. 
\begin{table}[h]
\caption{\textbf{Measured fidelities.} To quantify the propagation effect, we use the definition of fidelity $F(p,r) = \sum_i (p_i r_i)^{1/2}$, where $p$ and $r$ are the experimental and the theoretical discrete probability distributions of the different set of states with elements $p_i$ and $r_i$.}
\centering
\vspace{2mm}
\begin{tabular}{p{3cm} p{2cm} p{2cm}}
\toprule
\rule{0pt}{3ex}Set states &  &Fidelity ($\%$)\\[1ex]
\hline
\rule{0pt}{3ex}$\quad\;\ket{\psi_i}$&  &$94.58 \pm 0.02$\\[1ex]
$\quad\;\ket{\xi_j}$&  &$93.48 \pm 0.07$\\[1ex]
$\quad\;\ket{\varphi_m}$&  &$93.03 \pm 0.12$\\[1ex]
$\quad\;\ket{\phi_k}$&  &$90.81 \pm 0.02$\\[1ex]
\bottomrule
\bottomrule
\end{tabular}
\label{table:fidelity}
\end{table}
Figure~\ref{fig:tomos} shows the probability distribution matrices for the four different set of states, $\ket{\psi_i}$, $\ket{\xi_j}$, $\ket{\varphi_m}$ and $\ket{\phi_k}$, measured with mean photon number $\mu=0.011 \pm 0.002$ photon/pulse. In table~\ref{table:fidelity} the obtained fidelities are reported. The measured fidelity of the four different OAM states is $94.6\%$, and the transmission of superposition states between two ($\ket{\xi_j}$ and $\ket{\varphi_m}$) and four ($\ket{\phi_k}$) OAM modes leads to fidelities all larger than $90\%$, proving the robust conservation of all the phase relations after 1.2-km transmission, despite bends and twists of the fiber.
These results attest the feasibility of high-dimensional quantum communication with qudits encoded in their orbital angular momentum and represent a major step for high-dimensional quantum states distribution. Investigations on high-dimensional entangled states distribution for nonlocality tests, communication complexity problems and improved remote sensing are only a few of the foreseeable applications they can lead to.

\section{Orbital angular momentum based quantum key distribution}
\begin{figure*}[ht!]
     \centering
     \includegraphics[width=1\linewidth]{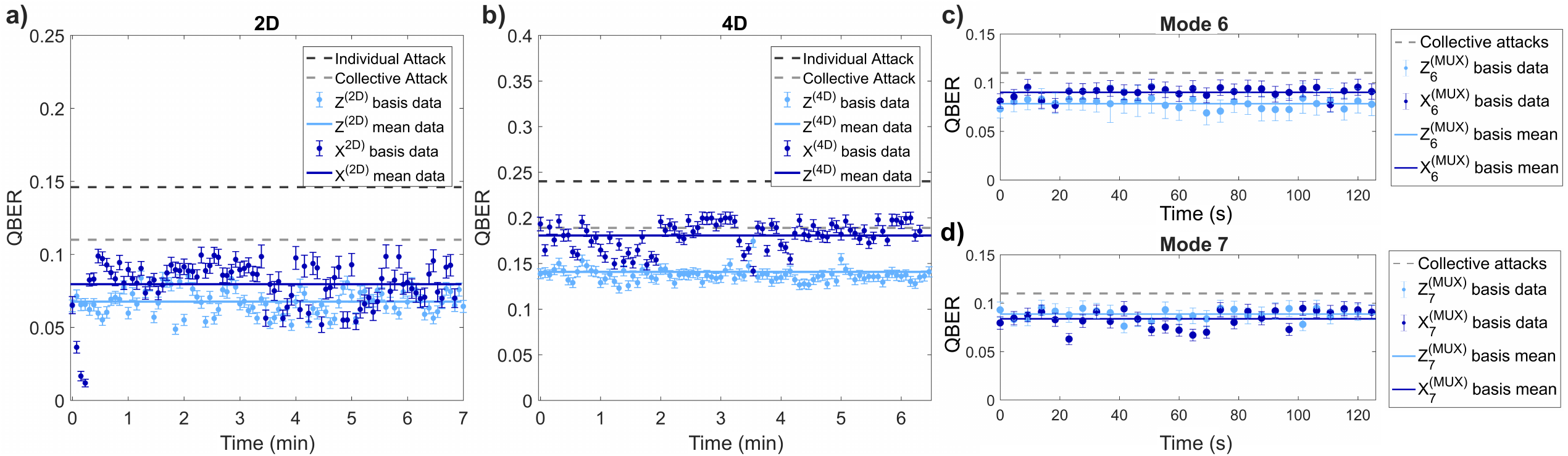}
   \caption{ {\bf QBERs as a function of time.}. Light blue dots are the experimental data collected in the computational bases and blue dots represent data for the Fourier bases. \textbf{a)} The average QBERs in the $\mathcal{Z}^{(2D)}$ and $\mathcal{X}^{(2D)}$ bases are $6.7$\% and $7.9$\% respectively. Both QBER values are below the 2$D$ individual and collective attacks limit, \textit{i.e.} $14.6\%$ and $11.0\%$ respectively. \textbf{b)}
   The average QBERs in the $\mathcal{Z}^{(4D)}$ and $\mathcal{X}^{(4D)}$ bases are $14.1$\% and $18.1$\% respectively. Both QBER values are below the 4$D$ individual and collective attacks limit, \textit{i.e.} $24.0\%$ and $18.9\%$ respectively. \textbf{c)} The average QBERs in the $\mathcal{Z}^{(MUX)}_6$ and $\mathcal{X}^{(MUX)}_6$ bases are $7.8$\% and $9.0$\% respectively. \textbf{d)} The average QBERs in the $\mathcal{Z}^{(MUX)}_7$ and $\mathcal{X}^{(MUX)}_7$ bases are $8.8$\% and $8.3$\% respectively. All error bars are obtained by considering a Poissonian statistic.}
\label{fig:QBERs}
\end{figure*}

\begin{figure}[h]
   \centering
    {\includegraphics[width=7cm]{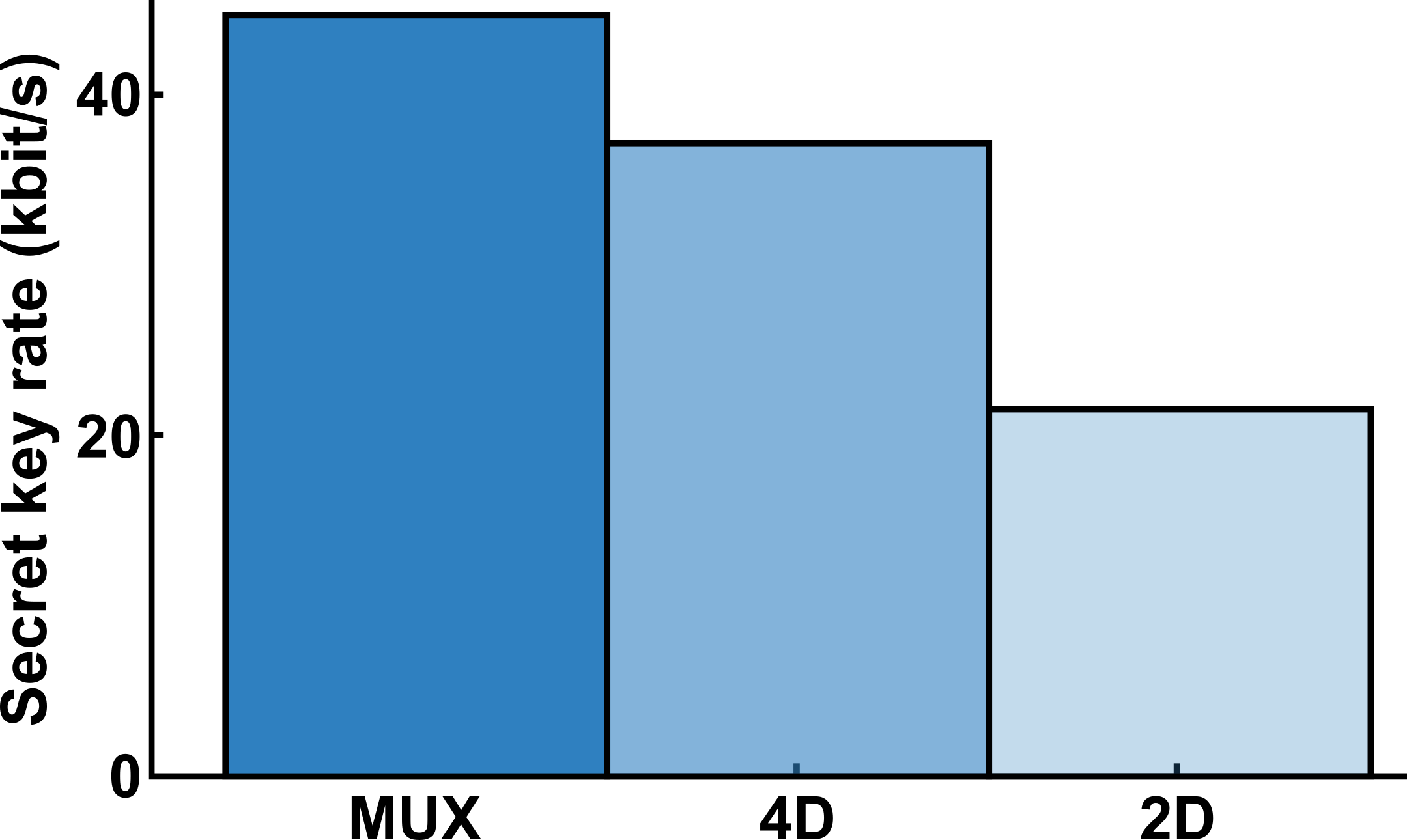}}
    \caption{\textbf{Secret key-rate comparison.} A comparison between the secret key rates achieved in the three QKD protocols is reported. The secret key rate for the 2D protocol is $22.81$ kbit/s, and $37.85$ kbit/s for the 4D case, marking an enhancement of the $69\%$ on the two-dimensional scheme. The final secret key rate for the multiplexed BB84 protocol (MUX) is $42.3$ kbit/s showing that multiplexing yields the highest key rates with our system.}
\label{fig:rsk}
\end{figure}

A branch of quantum communication that undoubtedly benefits from the capability of transmitting qudits is quantum key distribution~\cite{IslamTimeBin2017}. Indeed, having more information rate per transmitted photon, high-dimensional quantum states result in increased robustness to channel noise and higher key rates~\cite{Cerf2002,Durt2004,Erhard2018}. Hence, to show the practicality descending from the previous results and to characterize the OAM quantum communication system more rigorously, we perform three QKD protocols, which involve quantitative measures of the quantum bit error rate (QBER) of the transmitted quantum keys. We implemented 2$D$, 4$D$ and 2$\times$2$D$ multiplexed BB84 QKD protocols~\cite{BB84} using the experimental setup shown in figure~\ref{fig:mainsetup}. The optical switch allows the real-time preparation of the states within each mutually unbiased basis (MUBs) needed for the protocols. The two MUBs used for the 2D protocol are constituted by the states $\ket{\psi_2}$ and $\ket{\psi_4}$ for the computational bases $\mathcal{Z}^{(2D)}$, and the superpositions $\ket{\varphi_3}$ and $\ket{\varphi_4}$ for the Fourier bases $\mathcal{X}^{(2D)}$. The QBERs for both bases are measured, achieving average values of $6.7\%$ and $7.9\%$ respectively, which are below the two-dimensional collective attack threshold value of $11\%$~\cite{Cerf2002}. In Figure~\ref{fig:QBERs} a), the measured QBERs as a function of time are reported. The average QBER values lead to a positive secret key rate of $22.81$ kbit/s. In the 4D protocol case, the computational bases $\mathcal{Z}^{(4D)}$ is constituted by the entire set of $\ket{\psi_i}$, whereas the Fourier bases $\mathcal{X}^{(4D)}$ by the $\ket{\phi_k}$ states. The average QBERs measured are $14.1\%$ and $18.1\%$ respectively, which are lower than the four-dimensional collective attack threshold value of $18.9\%$~\cite{Cerf2002}. Figure~\ref{fig:QBERs} b) shows the measured QBERs as a function of time. The secret key rate extracted is $37.85$ kbit/s, proving an enhancement of the $69\%$ if compared with the 2D final key rate. This result confirms high-dimensional encoding as a superior scheme to achieve higher secret key rates in quantum key distribution~\cite{IslamTimeBin2017,cozzolinoPDP}. The parallel 2$\times$2$D$ multiplexed BB84 scheme is implemented using the $\ell=|6|$ and $\ell=|7|$ states separately for their own separate 2$D$ keys. The states $\ket{\psi_1}$, $\ket{\psi_2}$  and their superpositions $\ket{\xi_1}$, $\ket{\xi_2}$ constitute the basis $\mathcal{Z}^{(MUX)}_6$ and $\mathcal{X}^{(MUX)}_6$ for one key, while the states $\ket{\psi_3}$, $\ket{\psi_4}$ and their superpositions $\ket{\xi_3}$, $\ket{\xi_4}$ constitute the basis $\mathcal{Z}^{(MUX)}_7$ and $\mathcal{X}^{(MUX)}_7$ for the other key. To implement the multiplexed protocol, the experimental setup in figure~\ref{fig:mainsetup} is slightly adapted. The optical switch is substituted with a beam-splitter to simultaneously encode the keys on the two $|\ell|$ OAM quantum numbers. The average QBERs for the $\ell=|6|$ protocol are $7.8\%$ and $9.0\%$, producing a secret key rate of $21.58$ kbit/s. Conversely, average QBERs for the $\ell=|7|$ protocol are $8.8\%$ and $8.3\%$, producing a secret key rate of $20.72$ kbit/s. Thus, combining the two rates, a final secret key rate of $42.3$ kbit/s is achieved, \textit{i.e.} a higher key rate than for the 4$D$ case. Figures~\ref{fig:QBERs} c) and d) show the measured QBERs for the two parallel keys as function of time and Figure~\ref{fig:rsk} summarizes the comparison between the three QKD protocols. Further details on the QKD schemes implemented are reported in Appendices D and E.\\
In this work, we experimentally investigate the feasibility of using orbital angular momentum states in optical fiber as a vessel for transmitting high-dimensional quantum states. A reliable transmission of four different OAM quantum states and their superpositions is achieved and attested by fidelities above $90\%$. Owing to the robustness of the system, we implement three different QKD protocols, 2$D$, 4$D$ and 2$\times$2$D$. A 69\% key-rate enhancement of the 4$D$ over the 2$D$ scheme is achieved, resulting in the highest key rate of any high-dimensional QKD scheme based on spatial encoding to date.
These results show that high-dimensional quantum states can reliably and practically be transferred between two or more parties over km-length of the fiber in the telecom C band, entailing all the advantages deriving from higher-dimensional Hilbert spaces in quantum information and disclosing new scenarios for future quantum networks.

\section{Acknowledgments}
We thank C.~C.~W. Lim, S. Paesani, J. Cotler, D.~G. Marangon, P. Gregg and G. Carvacho for the fruitful discussions.
This work is supported by the Center of Excellence, SPOC-Silicon Photonics for Optical Communications (ref DNRF123), by the People Programme (Marie Curie Actions) of the European Union's Seventh Framework Programme (FP7/2007-2013) under REA grant agreement n$^\circ$ $609405$ (COFUNDPostdocDTU) and by the Office of Naval Research MURI program (N00014-13-1-0627) and the National Science Foundation (ECCS-1610190).

\appendix
\section{APPENDIX A: PREPARATION OF THE QUANTUM STATES}
The high-dimensional quantum states implemented in this work have OAM quantum number $\ell\in\lbrace -7,-6,+6,+7\rbrace$. They are prepared with two vortex plates (WPV10-1550, Thorlabs), which can be considered as nontunable q plates~\cite{qplate}. The vortex plates have topological charge $q_1=+3$ and $q_2=+7/2$ allowing the generation of the OAM states $\ket{+6},\ket{-6},\ket{+7}$ and $\ket{-7}$. Indeed, an OAM with different sign is added to the incoming photon depending on its spin angular momentum, \textit{i.e.} on its circular polarization handedness. The unitary transformation describing such a process is:
\begin{equation}
\begin{matrix}
  \ket{L,0} \xrightarrow{\pm q}\ket{R,\pm 2q}\\
  \ket{R,0} \xrightarrow{\pm q}\ket{L,\mp 2q}\\
\end{matrix}
\label{antial}
\end{equation}

So, the polarization states $\lbrace \ket{L}, \ket{R}\rbrace$ (left- and right-handed circular polarization) are adopted as standard basis and the diagonal and antidiagonal states, $\lbrace \ket{D}, \ket{A}\rbrace$, can be written as $\ket{D}=(\ket{R}+\ket{L})/\sqrt{2}$ and $\ket{A}=(\ket{R}-\ket{L})/\sqrt{2}$ respectively. Hence, a superposition among OAM states with the same $|\ell|$, is obtained by letting photons, with polarization states $\ket{D}$ or $\ket{A}$, pass through the vortex plate corresponding to that $|\ell|$. The superposition of quantum states with different $|\ell|$ is obtained by combining the states $\ket{|6|}$ and $\ket{|7|}$ with beam-splitters. Table~\ref{table:states} summarizes the quantum states and how they have been implemented.\\
The spin angular momentum ($\sigma^\pm$) is related to the left (right) -handed circular polarization by the following relations:

\begin{equation}
\begin{matrix}
   L \rightarrow \sigma^+\,,\quad \sigma^+=+\hbar \, ;\\
   R \rightarrow \sigma^-\,,\quad \sigma^-=-\hbar \, .\\
 \end{matrix}
\end{equation}

Thus, from equation \eqref{antial}, it is possible to recognize that, at the output of a vortex plate, OAM and SAM can be aligned or antialigned, that is OAM and SAM have the same or different sign. This property is fundamental for the transmission of the prepared quantum states through the fiber: aligned (antialigned) states are degenerate with each other (\textit{i.e.}, same effective index in the OAM fiber), but aligned states are not degenerate with the antialigned ones~\cite{Gregg2015}. In our experiment, we have chosen to work with antialigned states.

\section{APPENDIX B: AIR-CORE FIBER CHARACTERIZATION}
\begin{figure}[h]
  \centering
  \begin{minipage}[b]{0.46\textwidth}
    \includegraphics[width=\textwidth]{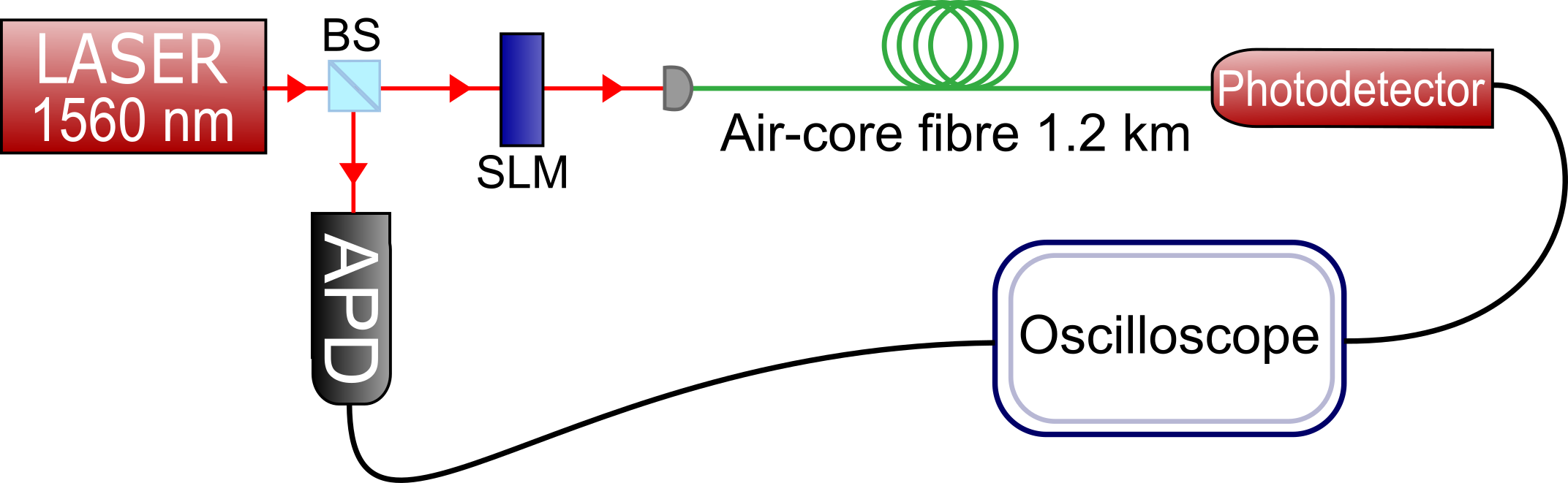}
    \caption{Setup of the TOA measurement. A pulsed laser at $1560$ nm is divided by a beam-splitter. One path is detected by a free-space avalanche photodiode and works as a trigger signal for the system. The other path is used to prepare different OAM modes ($\ell = \pm 5, \pm 6, \pm 7$) by using a spatial light modulator. Light carrying OAM is then transmitted over the $1.2$ km fiber. By using a $5$-GHz bandwidth In-Ga-As photodetector, it is possible to measure the different time of arrival of multiple modes.}
    \label{fig:TOA}
  \end{minipage}
  \hfill
  \begin{minipage}[b]{0.48\textwidth}
    \includegraphics[width=\textwidth]{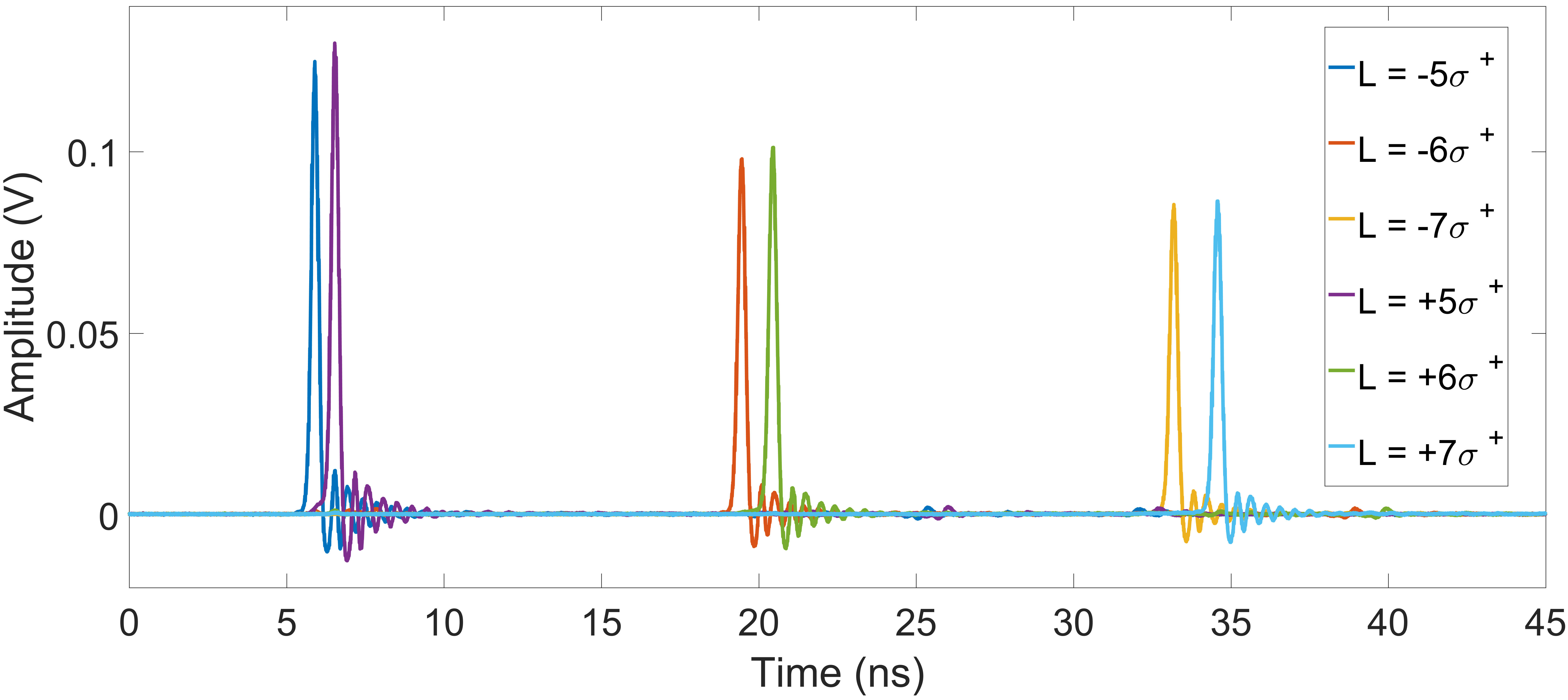}
    \caption{Time-domain plot of the multiple OAM modes, both for the antialigned and aligned spin-orbit condition. The antialigned modes are traveling faster than the aligned modes. The same happens to the modes with a low value of OAM quantum number with respect to those with a high value of $\ell$. In our setup, we compensate the time difference between the modes with $\ell=|6|$ and $\ell=|7|$ by designing a specific free-space time delay (approximately $4.5$ m) in order to receive the two modes at the same time.}
    \label{fig:modes}
  \end{minipage}
\end{figure}

\begin{figure}[]
  \centering
  \begin{minipage}[b]{0.46\textwidth}
    \includegraphics[width=\textwidth]{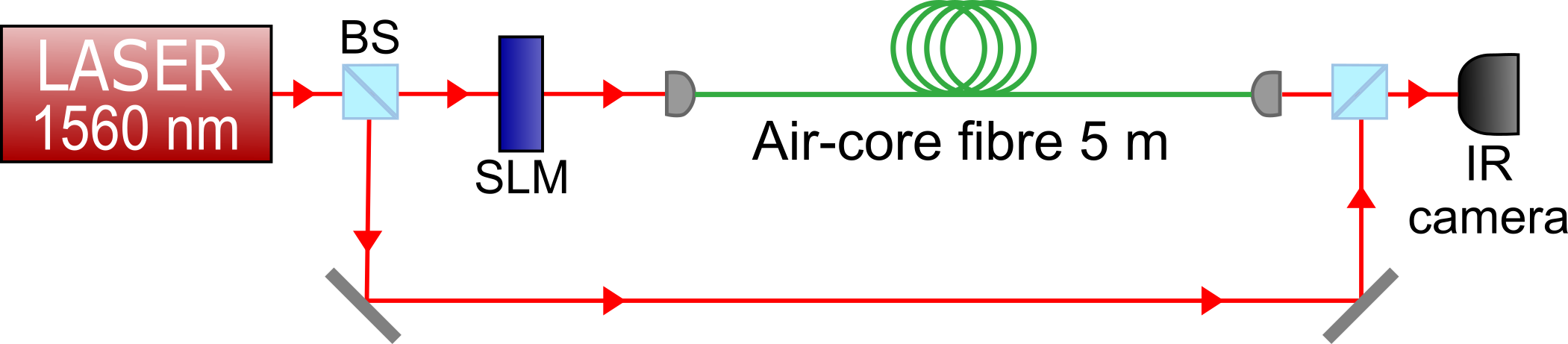}
    \caption{Setup for the OAM characterization. A pulsed laser at $1560$ nm is divided by a beam-splitter. One path is used to prepare different OAM modes ($\ell = \pm 5, \pm 6, \pm 7$), by using a SLM, which are sent through the air-core fiber. In the other path a Gaussian mode is directed to the fiber output so that it can interfere with the OAM beam. The interferograms are acquired with an infrared camera Xenics Xeva-4170.}
    \label{fig:OAM_Gauss}
  \end{minipage}
  \hfill
  \vspace{2mm}
  \begin{minipage}[b]{0.46\textwidth}
    \includegraphics[width=\textwidth]{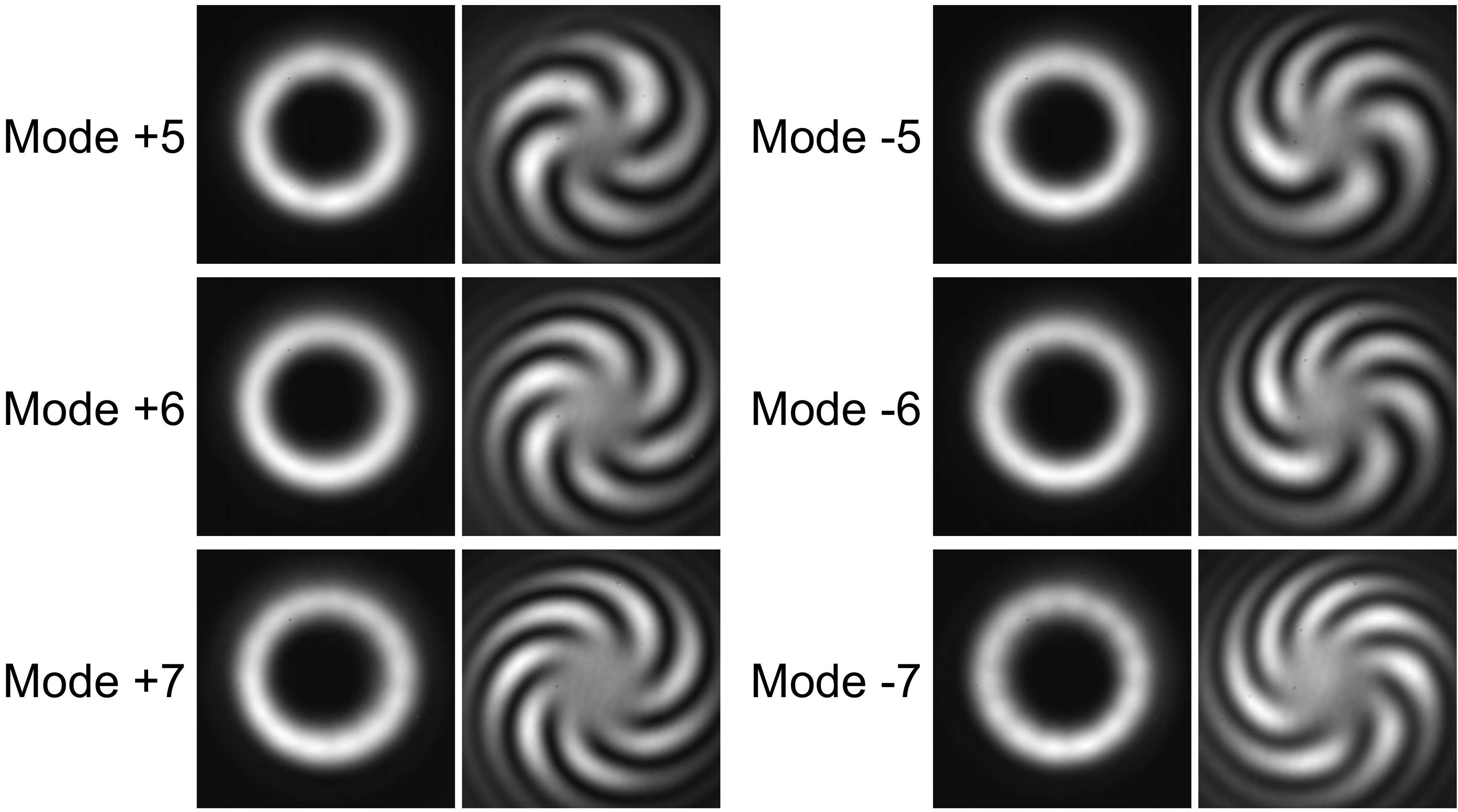}
    \caption{Images acquired by an infrared camera (Xenics Xeva-4170) of the various OAM modes ($\ell = \pm 5, \pm 6, \pm 7$) and their interference pattern with a Gaussian beam.}
     \label{fig:OAMinterf}
  \end{minipage}
\end{figure}

The fiber used enables conservation of the OAM modes, that is the information encoded in the OAM degree of freedom. An air core surrounded by a high-refractive index ring creates a large index difference between the center of the fiber and its outer part~\cite{Gregg2015}. This structure splits the spin-orbit-aligned (SAM and OAM with same sign) and spin-orbit-antialigned (SAM and OAM with opposed sign) states in effective index, thus allowing transmission of OAM modes with preserved orthogonality. The mode delay between $\ell=|6|$ and $\ell=|7|$, due to the different group velocities, is approximately $15$ ns and their extinction ratio (ER), $10\mathrm{log}(P_{max}/P_{min})$, is $18.4$ dB and $18.7$ dB respectively. A high extinction ratio is translated into a high purity of the modes and a low crosstalk between them. The measured ERs are higher than 18 dB for both cases, enabling faithful transmission of the states in the fiber. Note that crosstalk stemming from the coupling optics adds to the intrinsic fiber crosstalk. The fiber loss is around $1$ dB/km at $1550$ nm. To characterize the fiber we test the propagation of various OAM modes, \textit{i.e.} $\ell = \pm 5, \pm 6, \pm 7$, by using a $1560$ nm pulsed laser ($3$-ps pulse-width FWHM) and measuring their time of arrival (TOA). 
The setup used is depicted in Figure~\ref{fig:TOA}. The $1560$-nm pulsed laser is divided by a beam-splitter: one path is detected by an avalanche photodiode and works as a trigger signal for the oscilloscope, while the other is used to prepare the different OAM modes using a spatial light modulator (SLM).
\begin{table}[h!]
\caption{\textbf{Extinction ratio (dB) of the modes.} Measured extinction ratio between the modes $\ell=|5|, |6|, |7|$ with SAM and OAM antialigned. The fiber internal crosstalk and the free-space mode coupling affect the suppression between different modes.} 
\centering
\vspace{2mm}
\begin{tabular}{p{3cm} p{1.5cm} p{1.5cm} p{1.5cm}}
\toprule
\rule{0pt}{2.3ex}Detected mode&$\ell=|5|$&$\ell=|6|$&$\ell=|7|$\\ 
\hline
\rule{0pt}{2.3ex}ER&17.8&18.4&18.7\\ 
\bottomrule
\bottomrule
\end{tabular}
\label{table:SMcross}
\end{table}
\noindent
The beam carrying an OAM mode is then transmitted over the $1.2$-km fiber. By using a $5$ GHz bandwidth In-Ga-As free-space photodetector, it is possible to measure the different time of arrival of the multiple modes~\cite{Gregg2016}. The results of these measurements are shown in Figure~\ref{fig:modes}. A TOA measurement can be considered as a mode-purity indicator: a high extinction ratio of a mode with respect to the others means a high purity of the mode itself. In Table~\ref{table:SMcross}, the extinction ratios for each mode analyzed are reported. The special design of the air-core fiber allows the degeneracy of the antialigned (aligned) modes with each other. However, for order modes lower than $\ell=|4|$, this separation is not sufficient to allow transmission without crosstalk. For this reason the characterization is done with the modes $\ell = \pm 5, \pm 6\, \pm 7$. Since the in-fiber crosstalk is lower for the modes $\ell =|6|$ and $\ell =|7|$, we decide to implement the experiment with these. To prove that the modes generated own an orbital angular momentum, we make them interfere with a Gaussian beam. The setup used for the purpose and the interferograms acquired are shown in Figure~\ref{fig:OAM_Gauss} and
Figure~\ref{fig:OAMinterf}.

\section{APPENDIX C: EXPERIMENTAL SETUP}
 As shown in figure~\ref{fig:mainsetup}, an intensity modulator, controlled by a field programmable gate array, carves out pulses with a $600$-MHz repetition rate and a $150$-ps pulse width from a continuous-wave laser emitting at $1550$ nm. A variable optical attenuator is used to reach the quantum regime (mean photon number value: $\mu=0.011 \pm 0.002$ photon/pulse). A fast optical switch, driven by the same FPGA, allows the selection of the quantum states to be created between either the $\ket{\psi_i}$, $\ket{\xi_j}$, $\ket{\varphi_m}$ or $\ket{\phi_k}$. Phase modulators controlled by the FPGA determine the polarization of the weak pulses among diagonal or antidiagonal. Indeed, using polarization controllers (PCs) to adjust the input polarization to diagonal, we obtain the two outputs by applying voltages corresponding to 0 or $\pi$ phase shifts. Sets of half- and quarter-wave plates are used to convert from linear to circular polarization to create the states $\ket{\psi_i}$. The OAM is encoded in each photon by directing the photons through the two vortex plates Q1 and Q2. In particular, to obtain the superposition among different $\ket{|\ell|}$ states, linear optic devices are needed. As depicted in the scheme "States $\ket{\phi_k}$" embedded in figure~\ref{fig:mainsetup}, the pulses are split to follow two paths, each impinging on different vortex plates. A relative $\pi$ phase shift is added to change the phase relation between the two $|\ell|$ modes which are recombined. Before coupling the photons to the fiber, a mode delay to compensate the different group velocities between $\ell=|6|$ and $\ell=|7|$ in the air-core fiber is provided by optical delay lines. The quantum states are then coupled to the fiber with a $6$-mm lens mounted on a three-axis stage.\\
At the receiver side, an OAM mode sorter divides even from odd modes~\cite{Leach2002}. We first characterize the mode sorter with classical light, obtaining an extinction ratio of approximately $19$ dB. Furthermore, we use weak coherent pulses to measure the quantum visibility $V^{(|6|)}= 0.97\pm 0.03$ and $V^{(|7|)}= 0.98\pm 0.03$, where the visibility is defined as $V=(n_{max}-n_{min})/(n_{max}+n_{min})$ with $n_{max}$($n_{min}$) being the maximum (minimum) number of detected events. After the sorting process, the photons are converted back to the fundamental Gaussian-like mode by means of two vortex plates identical to Q1 and Q2. Projection measurements on the polarization infer the OAM quantum number sign. Two different schemes are implemented to measure the set of states $\ket{\psi_i}$, $\ket{\xi_j}$, $\ket{\varphi_m}$ or $\ket{\phi_k}$. In the case of $\ket{\varphi_m}$ and $\ket{\phi_k}$, a free-space Mach-Zehnder interferometer is cascaded with the mode sorter to measure the relative phase difference between the distinct $|\ell|$ modes. In one of the arms of the interferometer, half- and quarter-wave plates are used to compensate for polarization misalignments. The photons are detected by four In-Ga-As single-photon detectors (two ID230 and two ID220, ID Quantique), and registered by a time-tagger unit (ID801, ID Quantique). The insertion loss attributed to the receivers is measured to be approximately $9$ dB in receivers $\ket{\psi_i}$ and $\ket{\xi_j}$, and approximately $10$ dB in receivers $\ket{\varphi_m}$ and $\ket{\phi_k}$ (from the output of the OAM fiber to the input of the detectors). These losses can be further decreased by using noninterferometric mode sorters~\cite{ECOC2017}. Note that, the detection schemes implemented can register all the possible outcome events at the same time, so that complete device-independent demonstrations and loopholes-free measurements for nonlocality tests are possible~\cite{Weihs1198,Cyril2011}. Indeed, for a $D$-dimensional Hilbert-space-detection loophole-free test, $D+1$ outcomes are required to strictly violate Bell's inequalities. Conversely, projecting on $N<D$ outcomes, only a subset of all emitted pairs are measured, introducing possible classical correlation~\cite{Wang2018,kues2017}. 

\section{APPENDIX D: QUANTUM KEY DISTRIBUTION PROTOCOLS}
The protocols we implement can be considered as a BB84 with a three-intensity ($\mu$, $\nu$, $\omega$) decoy-state method, with dimensions $D=2$ and $D=4$. A second intensity modulator cascaded to that shown in figure~\ref{fig:mainsetup} implements the decoy-state technique by varying the amplitude of the weak coherent pulses. We indicate with $\mathcal{Z}$ the computational bases and with $\mathcal{X}$ the Fourier bases. The mutually unbiased bases used for the 2D protocol are:

\begin{equation}
\mathcal{Z}^{(2D)} = 
\begin{pmatrix}
     \ket{-6}\\
     \ket{-7}\\
\end{pmatrix} \,\quad \mathrm{and} \quad\,
\mathcal{X}^{(2D)} = \frac{1}{\sqrt{2}}
\begin{pmatrix}
     \ket{-6} + \ket{-7} \\
     \ket{-6} - \ket{-7} \\
\end{pmatrix} 
\end{equation}

The entire set of $\ket{\psi_i}$ and $\ket{\phi_k}$ states constitute the MUBs for the 4D protocol, that is:

\begin{equation}
\mathcal{Z}^{(4D)} =
\begin{pmatrix}
     \ket{+6}\\
     \ket{-6}\\
     \ket{+7}\\
     \ket{-7}
\end{pmatrix} 
\end{equation}
and
\begin{equation}
\mathcal{X}^{(4D)}=\frac{1}{2}
\begin{pmatrix}
     \ket{+6} + \ket{-6} + \ket{+7} + \ket{-7}\\
     \ket{+6} + \ket{-6} - \ket{+7} + \ket{-7}\\
     \ket{+6} - \ket{-6} + \ket{+7} - \ket{-7}\\
     \ket{+6} - \ket{-6} - \ket{+7} - \ket{-7}
\end{pmatrix} 
\end{equation}

In the case of the two multiplexed BB84, the MUBs for the $\ell=|6|$ and $\ell=|7|$ protocols are, respectively:

\begin{equation}
\mathcal{Z}^{(MUX)}_6 = 
\begin{pmatrix}
     \ket{+6}\\
     \ket{-6}\\
\end{pmatrix}  \,\quad \mathrm{and} \quad\,
\mathcal{X}^{(MUX)}_6 = \frac{1}{\sqrt{2}}
\begin{pmatrix}
     \ket{+6} + \ket{-6} \\
     \ket{+6} - \ket{-6} \\
\end{pmatrix}
\end{equation}

\begin{equation}
\mathcal{Z}^{(MUX)}_7 = 
\begin{pmatrix}
     \ket{+7}\\
     \ket{-7}\\
\end{pmatrix}  \,\quad \mathrm{and} \quad\,
\mathcal{X}^{(MUX)}_7 = \frac{1}{\sqrt{2}}
\begin{pmatrix}
     \ket{+7} + \ket{-7} \\
     \ket{+7} - \ket{-7} \\
\end{pmatrix}
\end{equation}

The secret key rate for a QKD protocol with decoy-states is~\cite{IslamTimeBin2017}:

\begin{equation}
R_{SK} = Q_{\mathcal{Z},1} [  \log_2 {D}-h_{D} \lbrace e_{\phi,1} \rbrace ]- Q_{\mathcal{Z}} \Delta_{leak}
\end{equation}

where $Q_{\mathcal{Z},1}$ is the single-photon gain defined as:

\begin{equation}
 Q_{\mathcal{Z},1} = [ p_\mu \mu e^\mu + p_\nu \nu e^\nu +p_\omega \omega e^\omega] \: Y_{\mathcal{Z},1} 
\end{equation}
with $\mu$, $\nu$, $\omega$ the values of the different intensities and $p_\mu,p_\nu,p_\omega$ the probabilities of choosing them, respectively. The intensities used for the three protocols are reported in Table~\ref{table:decoy}. 
\begin{table}[]
\caption{\textbf{Decoy-state intensities.} The table shows the mean photon number per pulse for each decoy-state intensity used in the three protocols.}
\centering
\begin{tabular}{p{2cm} p{2cm} p{1cm}}
\toprule
\rule{0pt}{0.3ex}$\;\;\;\mu$ & $\;\;\;\nu$ & $\;\;\;\omega$\\[0.3ex]
\hline
0.011 & 0.008 & 0.000\\[0.3ex]

\bottomrule
\bottomrule
\end{tabular}
\label{table:decoy}
\end{table}
The quantity $h_{D}$ is the mutual information in the $D$-dimensional case, defined as $h_{D}=-x \log_2 (x/(D-1))-(1-x) \log_2 (1-x)$. The single-photon yield in the $\mathcal{Z}$ bases is defined as

\begin{equation}
\begin{split}
Y_{\mathcal{Z},1}  = \max \Big \lbrace \frac{\mu}{\mu \nu - \mu \omega -\nu^2 +\omega^2 } \Big [ Q_{\mathcal{Z},\nu} e^\nu \\
- Q_{\mathcal{Z},\omega} e^\omega -\frac{\nu^2-\omega^2}{\mu^2} \Big ( Q_{\mathcal{Z},\mu} e^\mu - Y_{\mathcal{Z},0}  \Big) \Big ]  ,0 \Big \rbrace. 
\end{split}
\label{eq:gain_M0}
\end{equation}

$Y_{\mathcal{Z},0}$ is the zero-photon yield in $\mathcal{Z}$, bounded by:

\begin{equation}
Y_{\mathcal{Z},0}  = \max \Big \lbrace \frac{\nu Q_{\mathcal{Z},\omega} e^\omega - \omega Q_{\mathcal{Z},\nu} e^\nu}{\nu-\omega} ,0 \Big \rbrace. 
\end{equation}

Finally the single-photon error rate in the $\mathcal{X}$ bases can be written as:

\begin{equation}
e_{\mathcal{X},1}  = \min \Big \lbrace \frac{e_{\mathcal{X},\nu} \: Q_{\mathcal{Z},\nu} \: e^\nu - e_{\mathcal{X},\omega} \: Q_{\mathcal{Z},\omega} \: e^\omega}{(\nu-\omega) \: Y_{\mathcal{X},1}} ,0 \Big \rbrace. 
\end{equation}

where $e_{\mathcal{X},\nu}$ and $e_{\mathcal{X},\omega}$ are the error rate in the $\mathcal{X}$ bases for the $\nu$ and $\omega$ intensities. $Y_{\mathcal{X},1}$ can be estimated by using equation~\ref{eq:gain_M0} for the $\mathcal{X}$ bases. 
Note that in the implementation of the 4D QKD protocol, we use three out of four states of the $\mathcal{X}^{(4D)}$ basis: $\ket{\phi_1}$, $\ket{\phi_2}$ and $\ket{\phi_3}$. This does not limit the security of the system, as demonstrated in Refs.~\cite{fewer2018,Tamaki2014}, but it helps the preparation and detection of the $\ket{\phi_k}$ states, resulting in a lower QBER for the $\mathcal{X}^{(4D)}$ basis. The experimental parameters used to extract the secret key rate are reported in Table~\ref{table:par}. Here, we assume that the transmitter and the receiver exchange an infinitely long key for simplicity, although a finite-key scenario could equally be considered for the key extraction~\cite{Ma2005,fewer2018}.
\begin{table}[h]
\caption{\textbf{Experimental parameters for the secret key-rate extraction.}}
\centering
\vspace{2mm}
\begin{tabular}{c c c c c c c c}
\toprule
\rule{0pt}{0.3ex}$Q_\mu$&$Q_\nu$&$Q_\omega$&$p_{\mathcal{Z}}$&$p_{\mathcal{X}}$&$p_\mu$&$p_\nu$&$p_\omega$\\ [1.0ex] 
$1.6\times 10^{-4}$&$1.4\times 10^{-4}$&$3.2\times 10^{-7}$&$0.9$&$0.1$&$0.98$&$0.01$&$0.01$\\
\bottomrule
\bottomrule
\end{tabular}
\label{table:par}
\end{table}

\section{APPENDIX E: DETECTION PROBABILITIES OF THE QKD PROTOCOLS}
\begin{figure}[h]
  \centering
  \begin{minipage}[t]{0.44\textwidth}
    \includegraphics[width=\textwidth]{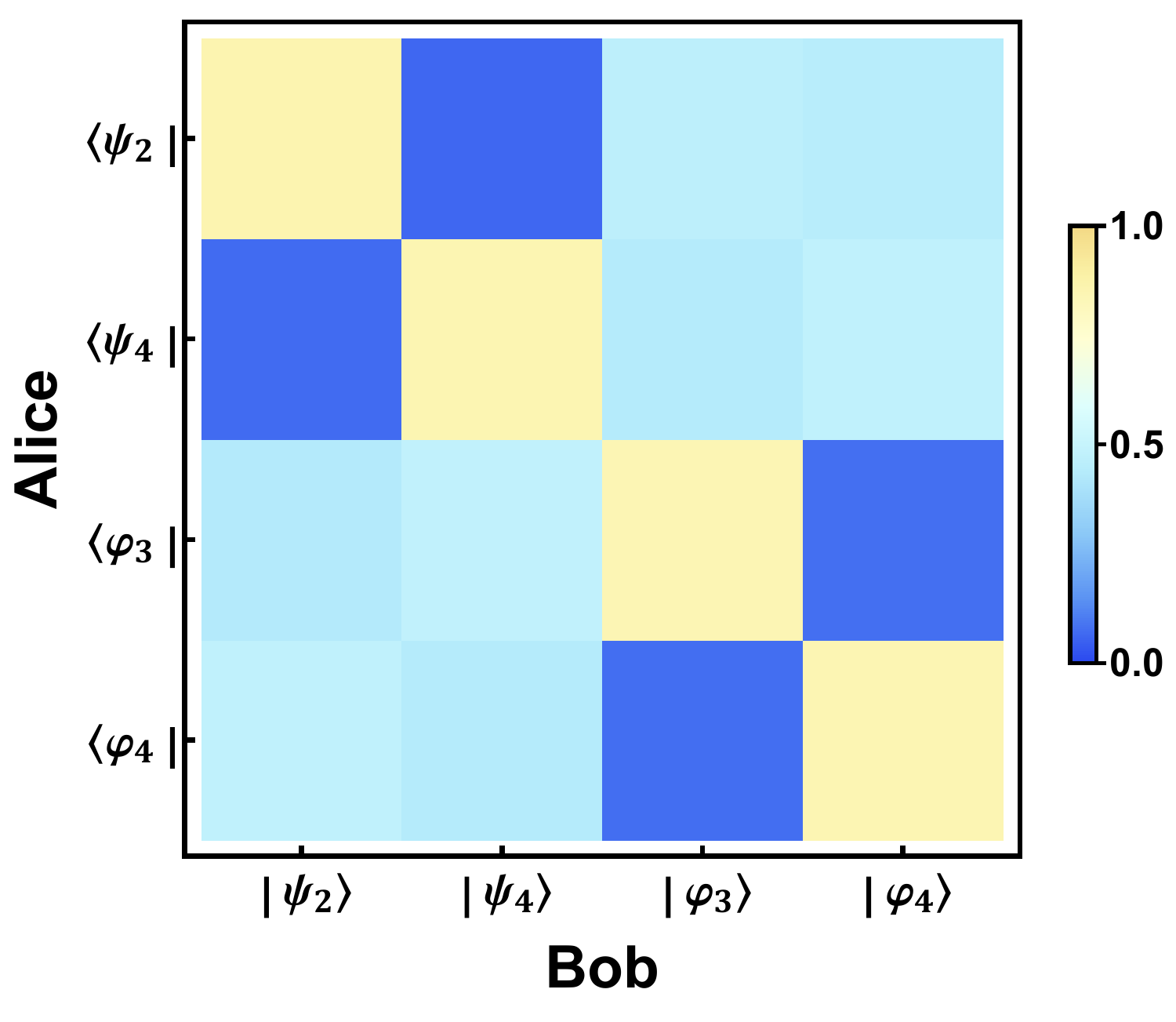}
    \caption{\textbf{2D protocol detection probability.} The average fidelity measured over $2$ minutes is $(98.0\pm 0.2)\%$.}
    \label{fig:tomo2d}
  \end{minipage}
  \hfill
  \begin{minipage}[t]{0.44\textwidth}
    \includegraphics[width=\textwidth]{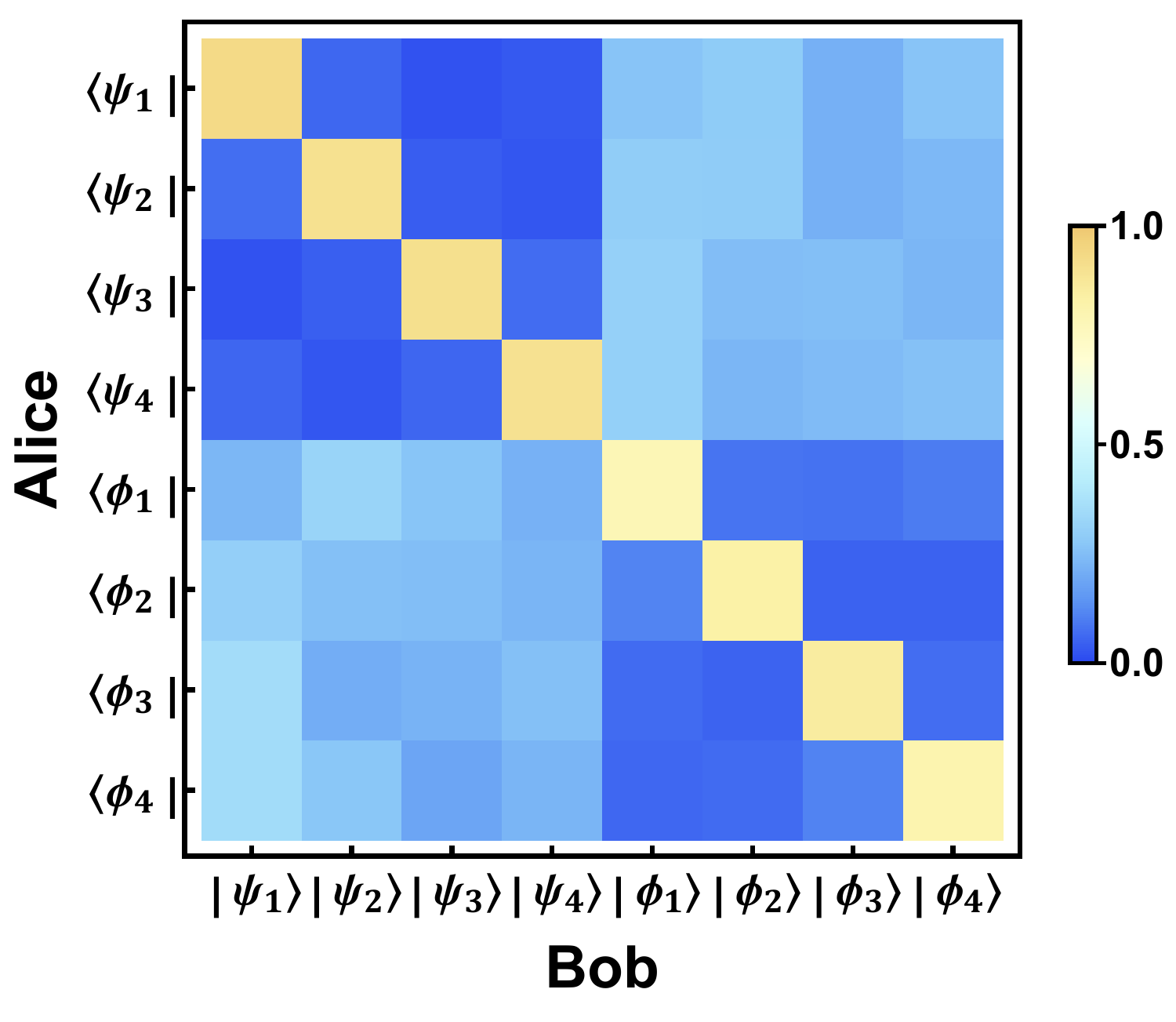}
    \caption{\textbf{4D protocol detection probability.} The average fidelity measured over $5$ minutes is $(95.8\pm 0.4)\%$.}
    \label{fig:tomo4d}
  \end{minipage}
\end{figure}

\begin{figure}[h]
  \centering
  \begin{minipage}[t]{0.44\textwidth}
    \includegraphics[width=\textwidth]{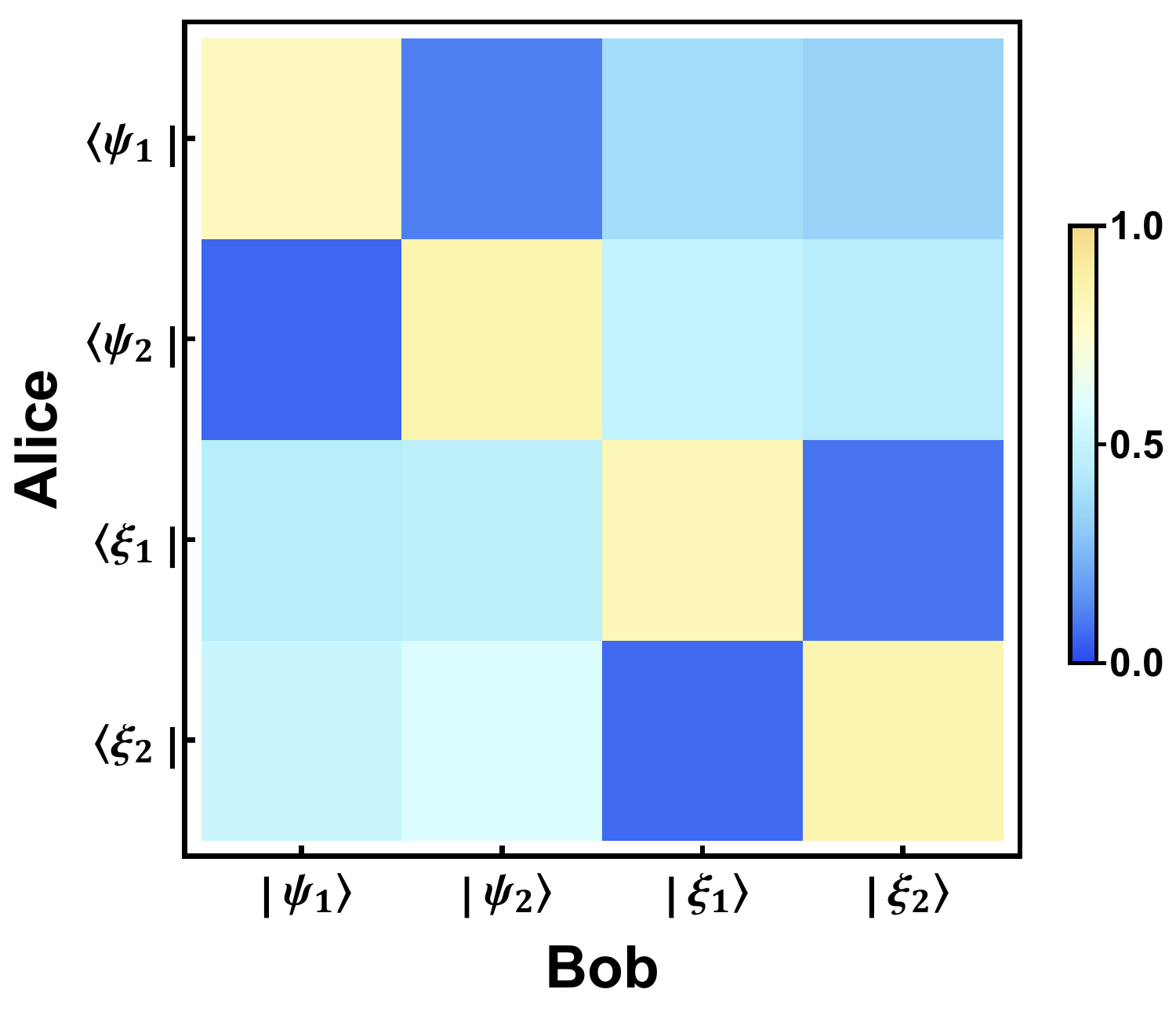}
    \caption{\textbf{Mode 6 multiplexed detection probability.} The average fidelity measured over $2$ minutes is $(96.6\pm 0.2)\%$.}
    \label{fig:mux6}
  \end{minipage}
  \hfill
  \begin{minipage}[t]{0.44\textwidth}
    \includegraphics[width=\textwidth]{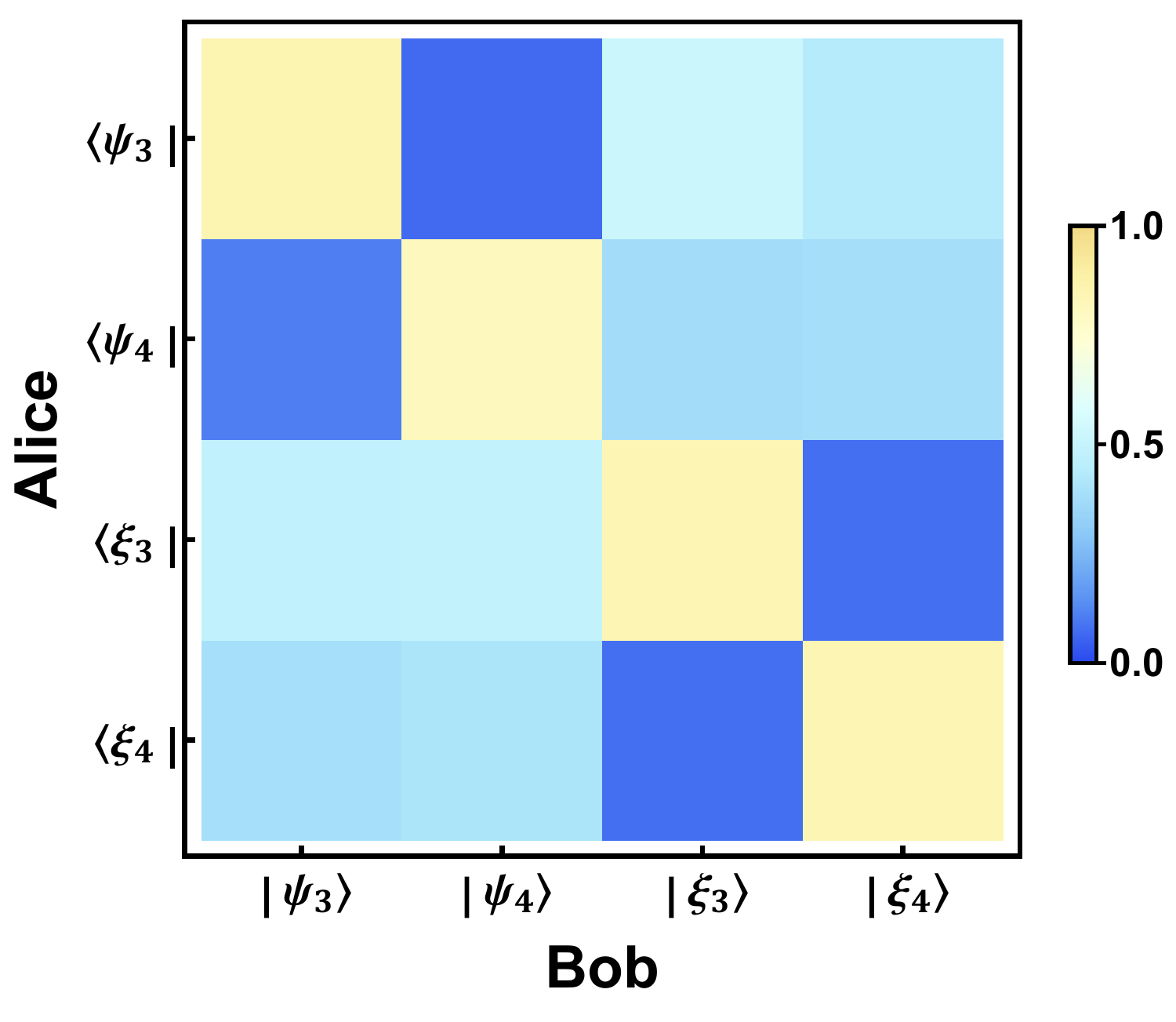}
    \caption{\textbf{Mode 7 multiplexed detection probability.} The average fidelity measured over $2$ minutes is $(96.7\pm 0.3)\%$.}
    \label{fig:mux7}
  \end{minipage}
\end{figure}

To characterize the setup, we measure the detection probability for all the quantum key distribution protocols implemented. Figure~\ref{fig:tomo2d} and~\ref{fig:tomo4d} show the two matrices for the 2$D$ and 4$D$ cases. The average fidelities obtained are $(98.0\pm 0.2)\%$ and $(95.8\pm 0.4)\%$, respectively. The same procedure is carried out for the multiplexed BB84 protocol. Figures~\ref{fig:mux6} and~\ref{fig:mux7} show the two matrices experimentally obtained, whose fidelities are $(96.6\pm 0.2)\%$ and $(96.7\pm 0.3)\%$ for the OAM quantum number $\ell=|6|$ and $\ell=|7|$ respectively.

\section{APPENDIX F: ELECTRONIC DESIGN}
\begin{figure}[h]
  \centering
  \begin{minipage}[t]{0.44\textwidth}
    \includegraphics[width=\textwidth]{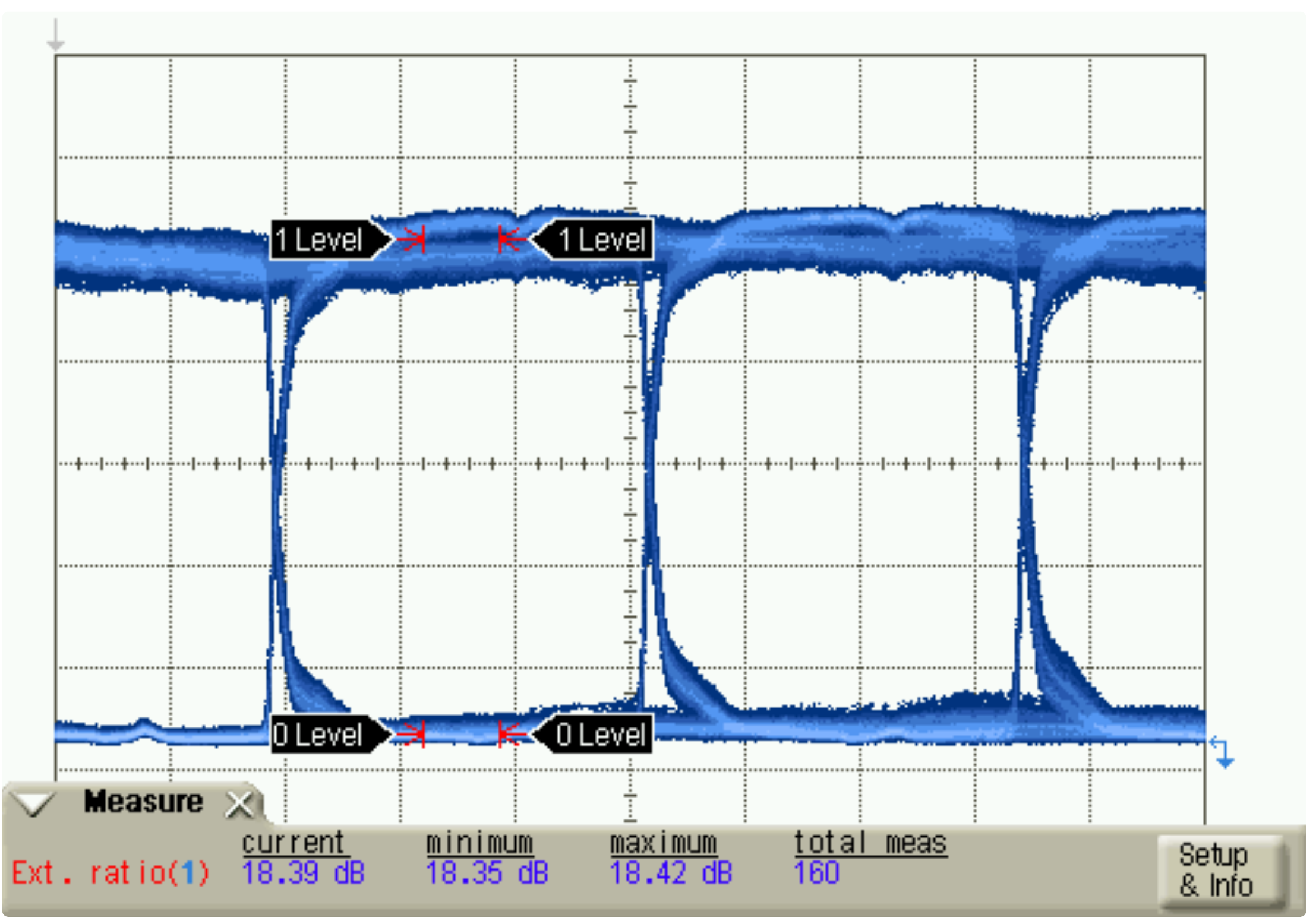}
    \caption{Eye diagram of the intensity modulator, iXblue MZ-LN-10. Extinction ratio $18.4$ dB}
    \label{fig:IM}
  \end{minipage}
  \hfill
  \begin{minipage}[t]{0.44\textwidth}
    \includegraphics[width=\textwidth]{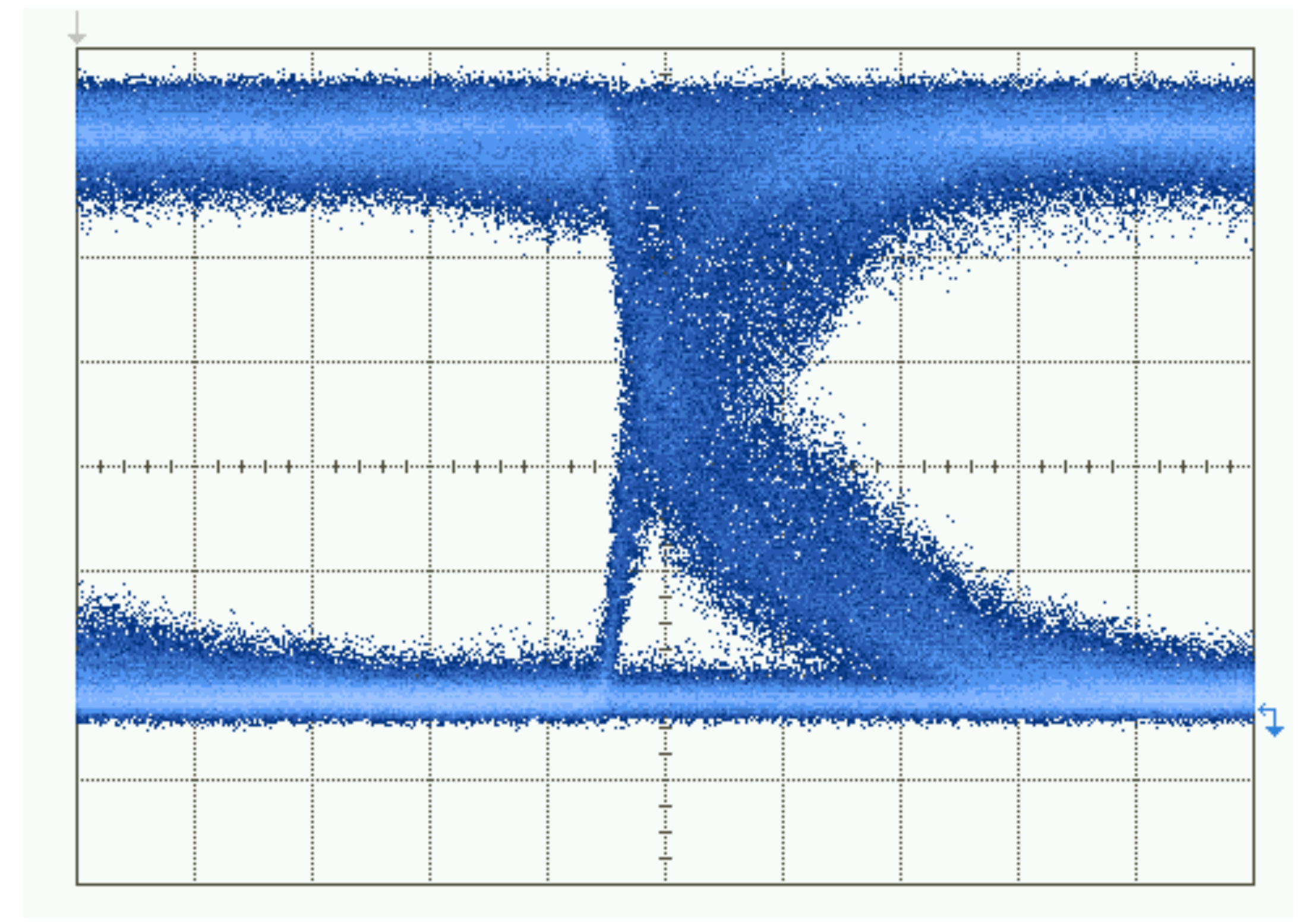}
    \caption{Eye diagram of the fast optical switch, EOspace AX-2x2-0K5-10-PFA-PFA. Extinction ratio $19.2$ dB.}
    \label{fig:MZI}
  \end{minipage}
\end{figure}

\begin{figure}[h]
  \centering
  \begin{minipage}[b]{0.44\textwidth}
    \includegraphics[width=\textwidth]{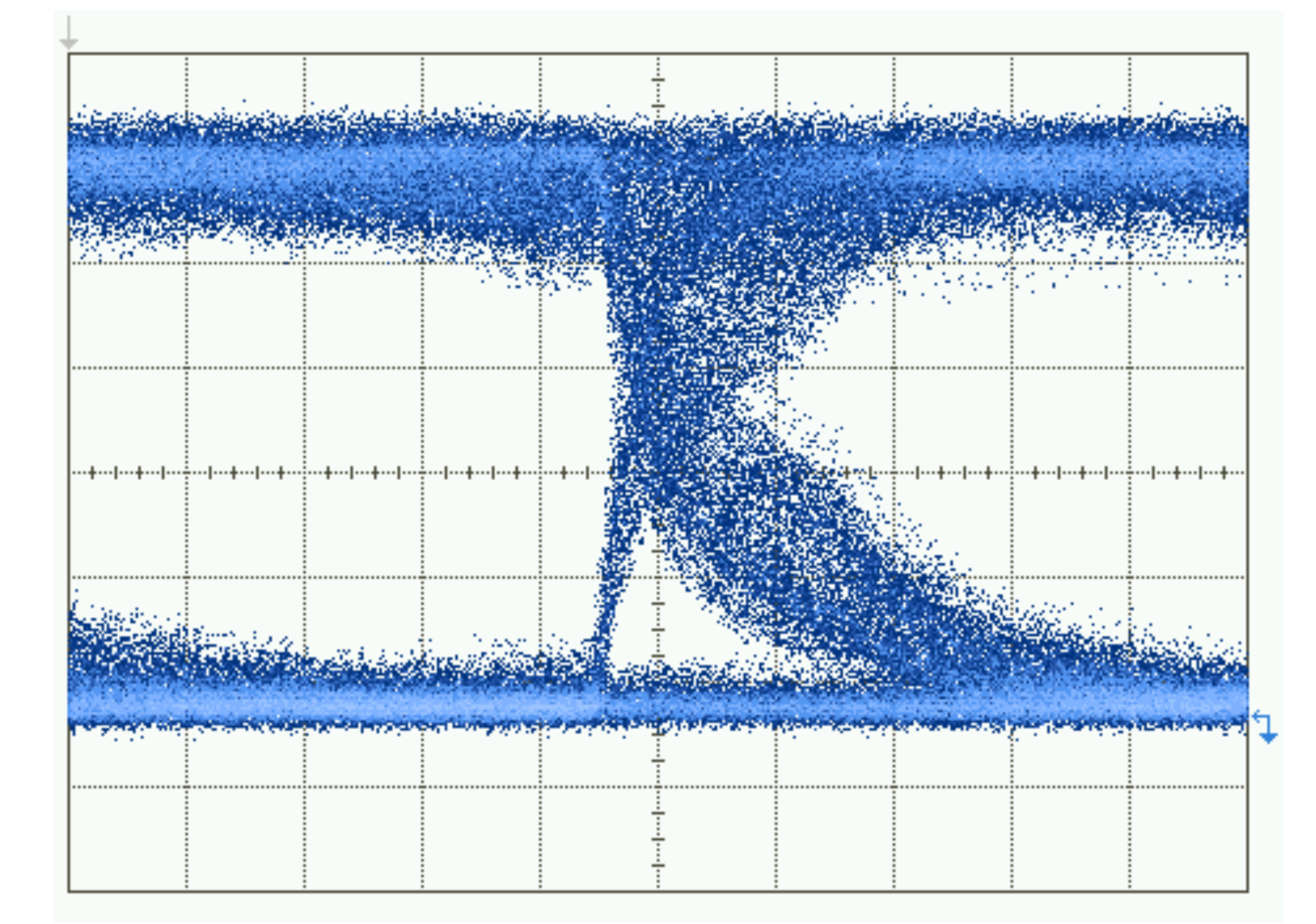}
    \caption{Eye diagram of the phase modulator, Photline MPZ-LN-10. Extinction ratio $17.6$ dB.}
    \label{fig:PM1}
  \end{minipage}
  \hfill
  \begin{minipage}[b]{0.44\textwidth}
    \includegraphics[width=\textwidth]{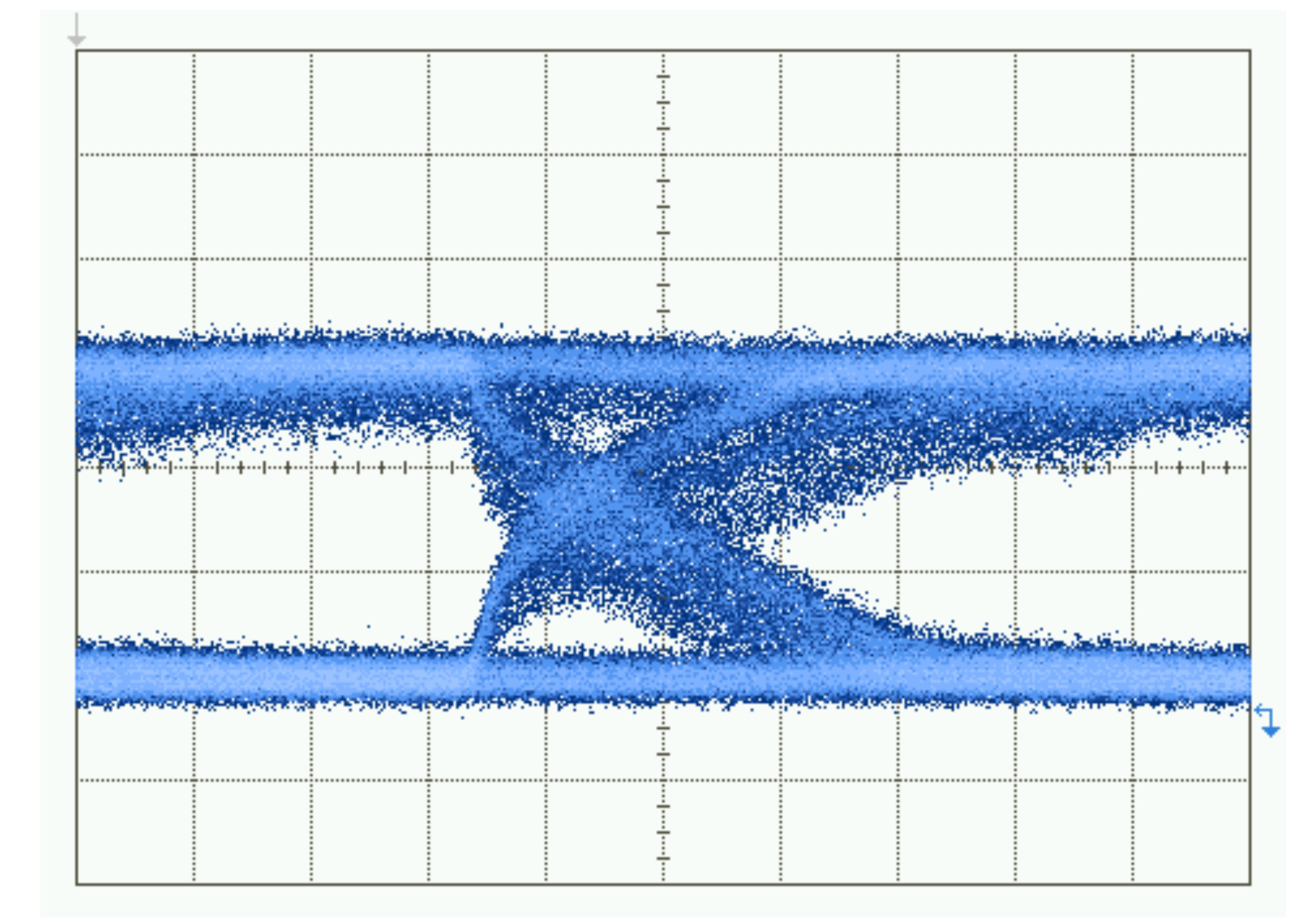}
    \caption{Eye diagram of the phase modulator, iXblue MPZ-LN-10. Extinction ratio $13.3$ dB.}
    \label{fig:PM2}
  \end{minipage}
\end{figure}

The opto-electronic devices used in the experiment have been characterized with classical light. The eye diagrams for the intensity modulator, the optical switch and the two phase modulators are taken from an oscilloscope and are shown in figures~\ref{fig:IM},~\ref{fig:MZI},~\ref{fig:PM1} and ~\ref{fig:PM2}. The FPGA generates five digital outputs: a train of pulses to drive the intensity modulator, three pseudorandom binary sequences (PRBSs) of $2^{12}-1$ bit (where a bit slot corresponds to a duration $T\approx 1.68$ ns) to drive the optical switch and the two phase modulators, and an electrical pulsed signal at $145$ kHz to synchronize the transmitter and receiver. The repetition rate is $600$ MHz, the electrical pulse width is approximately $100$ ps, whereas the optical pulse width is approximately $150$ ps.

\clearpage
\newpage


%

\end{document}